\documentclass[
11pt,%
tightenlines,%
nofootinbib,preprintnumbers,prd]{revtex4}
\usepackage{graphicx}
\usepackage{amssymb}
\newcommand{\be}{\begin{equation}}
\newcommand{\ee}{\end{equation}}
\newcommand{\bea}{\begin{eqnarray}}
\newcommand{\eea}{\end{eqnarray}}
\newcommand{\nn}{\nonumber}
\newcommand{\0}{\over }
\newcommand{\2}{{1\over2}}
\newcommand{\4}{{1\over4}}
\newcommand{\8}{{1\over8}}

\def\0{\over } \def\2{{1\over2}} \def\4{{1\over4}}
\def\5{\hat } \def\6{\partial }

\def\({\left(} \def\){\right)} \def\<{\langle } \def\>{\rangle }

\begin{document}
\preprint{SACLAY-T03/024}
\preprint{TUW-03-08}
\title{On the Apparent Convergence of Perturbative QCD at High Temperature
}
\author{J.-P. Blaizot}
\affiliation{Service de Physique Th\'eorique, CE Saclay,
         F-91191 Gif-sur-Yvette, France}
\author{E. Iancu}
\affiliation{Service de Physique Th\'eorique, CE Saclay,
         F-91191 Gif-sur-Yvette, France}
\author{A. Rebhan}
\affiliation{Institut f\"ur Theoretische Physik, Technische
Universit\"at Wien, 
A-1040 Vienna, Austria }
\begin{abstract}
The successive perturbative estimates of the
pressure of QCD at high temperature $T$  show no sign of 
convergence,  unless the coupling constant $g$ is unrealistically small.
Exploiting known results of an
effective field theory which separates hard (order $2\pi T$) and soft (order
$gT$) contributions, we explore the accuracy of simple resummations which
at a given loop order systematically treat hard contributions
strictly perturbatively, but soft contributions without truncations.
This turns out to improve significantly the two-loop and the three-loop results
in that both remain below the ideal-gas value, and the
degree of renormalization
scale dependence decreases as one goes from two to three loop order,
whereas it increases in the conventional perturbative results.
Including the four-loop logarithms recently obtained by Kajantie et al.,
we find that this trend continues and that with a particular 
sublogarithmic constant the untruncated four-loop
result is close to the three-loop result, which itself agrees
well with available lattice results down to
temperatures of about $2.5T_c$.
We also investigate the possibility of optimization
by using a variational (``screened'')
perturbation theory in the effective  theory.
At two loops, this gives a result below the
ideal gas value, and also closer to
lattice results than the recent two-loop hard-thermal-loop-screened result
of Andersen et al.
While at three-loop order the gap equation of
dimensionally reduced screened perturbation
theory does not have a solution in QCD, this is
remedied upon inclusion of the four-loop logarithms.
\end{abstract}
\maketitle

\section{Introduction}

\begin{figure}
\centerline{\includegraphics[bb=70 200 540 550,width=10cm]{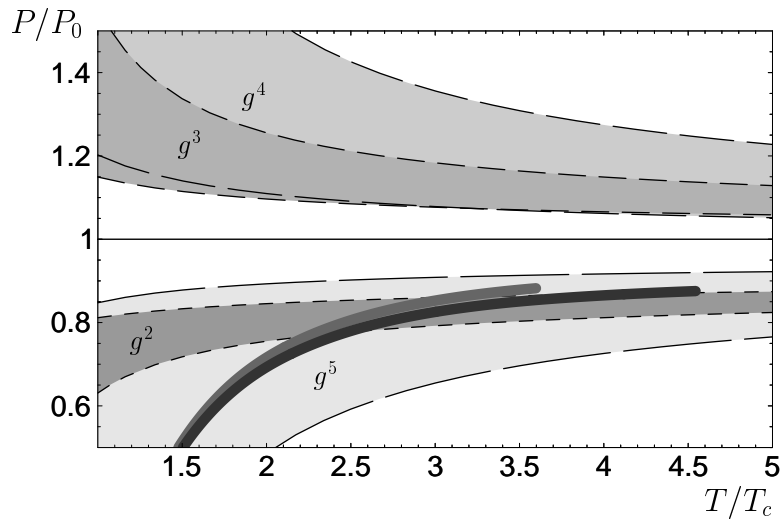}
}
\caption{Strictly perturbative results for the thermal pressure
of pure glue QCD normalized to the ideal-gas value, as a function
of $T/T_c$ (assuming $T_c/\Lambda_{\overline{\hbox{\scriptsize MS}}}=1.14$).
The various gray bands bounded by differently
dashed lines show the perturbative results
to order $g^2$, $g^3$, $g^4$, and $g^5$, with $\overline{\hbox{MS}}$
renormalization
point $\bar\mu$ varied between $\pi T$ and $4\pi T$. The thick
dark-grey line shows the continuum-extrapolated
lattice results from Ref.~\protect\cite{Boyd:1996bx};
the lighter one behind that of a lattice calculation
using an RG-improved action \cite{Okamoto:1999hi}.
\label{fig:qcd}}
\end{figure}

One could expect weak coupling calculations to lead to reasonable 
estimates of the QCD free
energy at  high temperature $T$, a regime where indeed
the gauge coupling becomes
small because of asymptotic freedom.
But explicit perturbative calculations, which have been pushed in 
recent years up to
the order $g^{5}$
\cite{Shuryak:1978ut,Kapusta:1979fh,Toimela:1983hv,Arnold:1995eb,Zhai:1995ac,Braaten:1996ju,Braaten:1996jr} 
do not exhibit
any sign of convergence, as depicted in Fig.~\ref{fig:qcd}; they 
rather show increasing ambiguities due
to the dependence on the renormalization point, signalling a complete
loss of predictive power.

Various mathematical extrapolation techniques  have been tried,
such as  Pad\'e approximants
\cite{Kastening:1997rg,
Cvetic:2002ju} and
Borel resummation \cite{Parwani:2000rr
}. The resulting expressions are
indeed  smooth functions of the coupling, better behaved than polynomial
approximations truncated at order $g^5$ or lower, with a weaker 
dependence on the
renormalization scale. However, while these 
methods do improve the situation somewhat,  it is fair to
say that they offer little physical insight on the source of the difficulty.

Recognizing that an important effect of thermal fluctuations is to 
give a mass to the excitations, thereby screening long
range interactions, it has been suggested to incorporate such 
screening effects in the tree-level Lagrangian, and correct
for double counting by adding suitable counterterms to the 
interactions. Such a scheme has been implemented with some success in
scalar field theory under the name of screened perturbation theory (SPT)
~\cite{Karsch:1997gj,Chiku:1998kd,Andersen:2000yj,Andersen:2001ez}.
It has been extended to QCD by including at tree level
the entire non-local Lagrangian of the hard thermal loops
\cite{Andersen:1999fw,Andersen:1999sf,Andersen:2002ey,Andersen:2003zk},
which is referred to as HTL perturbation theory (HTLPT).

A different approach, motivated physically by the success of the 
quasiparticle picture, is based on the so-called
$\Phi$-derivable approximations
\cite{Blaizot:1999ip,
Blaizot:2000fc,BIR:Review}
(see also Peshier \cite{Peshier:2000hx}). 
This approach takes advantage of remarkable 
simplifications which occur in the calculation of the
entropy at two loop in the skeleton expansion.
Together
with further approximations\footnote{A full $\Phi$-derivable
approximation has been worked out successfully in
scalar field theory~\cite{Braaten:2001en}, 
for which recently also the question
of renormalizability could be answered affirmatively
in Refs.~\cite{vanHees:2001ik,
Blaizot:2003br}.}
for the self-energies based
on hard thermal loops, this led to results for the thermodynamical
functions which are consistent with lattice calculations
for temperatures above 2.5 $T_c$.

This paper reconsiders the known results up to order $g^5$ in the light of
the  simple observation  that  the accuracy of perturbation theory is 
not the same at all
momentum scales. Thus, while perturbation theory at the scale $T$ is 
an expansion in powers of $g^2$, the perturbation
theory at the scale
$gT$ is an expansion in powers of $g$, and is therefore less 
accurate. It is  when they are treated strictly perturbatively
that  the soft contributions turn out to completely spoil apparent 
convergence, as exemplified by the (soft) contribution of
order
$g^3$   which leads to a pressure exceeding the ideal gas value. The 
main idea that we want to pursue is to decouple
approximations in the hard and the soft sectors: in the hard sector, 
we shall use perturbation theory since it is accurate;
in the soft sector we shall go (minimally) beyond perturbation theory,
with the remarkable result that the apparent convergence of
perturbative QCD at high temperatures is dramatically improved. This,
in our opinion, lends support to the more ambitious
attempts to reorganize perturbation theory by novel resummation techniques.

The tools to deal with various momentum scales in field theory are 
well developed and involve the construction of effective
theories. In fact, the perturbative results through order $g^5$,
first calculated in Refs.~\cite{Arnold:1995eb,Zhai:1995ac},
have been rederived and confirmed by Braaten and Nieto \cite{Braaten:1996jr}
by an elegant and efficient
effective-field-theory method
which separates contributions from hard, soft, and supersoft
momentum scales, $2\pi T$, $gT$, and $g^2T$, respectively.
In Euclidean space, the only soft modes are static ones,
and the effective field theories for these are therefore
dimensionally reduced \cite{Gross:1981br,Appelquist:1981vg,Nadkarni:1983kb} to
three-dimensional ones.
The dimensionally reduced theory consists of
massive adjoint scalar fields $A_0^a$ and massless
three-dimensional Yang-Mills fields (with
coupling $g_E$) and, following Braaten and Nieto, we shall refer to it as
  electrostatic QCD (EQCD) in the following.
The corresponding  effective
three-dimensional Lagrangian is
\be\label{LQCDdr}
\mathcal L_{EQCD}=\4 F_{ij}^aF_{ij}^a + \2 (D_i A_0)^a(D_i A_0)^a
+\2 m_E^2 A_0^aA_0^a + \8\lambda_E (A_0^aA_0^a)^2 +\delta \mathcal L_{EQCD}
\ee
where the parameters are determined perturbatively by matching
\cite{Braaten:1996jr}. In lowest order we have:
\be\label{LQCDparam}
m_E^2=(1+N_f/6)g^2 T^2,\quad\qquad g_E^2=g^2T,
\ee
and
\cite{Nadkarni:1988fh}
\be\label{lambdaELO}
\lambda_E={9-N_f\012\pi^2}g^4T.
\ee
In fact, $\lambda_E$ starts to contribute to the
pressure only at order $g^6$.
For this reason we shall ignore this particular coupling
in most of the following, as well as all the other vertices
contained in $\delta\mathcal L_{EQCD}$.

The thermal pressure of the 4-dimensional theory can be
decomposed into contributions from the hard modes $\sim T$,
calculable by standard perturbation theory, and soft contributions
governed by (\ref{LQCDdr}) which involves both perturbatively
calculable contributions up to order $g^5 T^4$ and
nonperturbative ones coming from the fact that the
effective theory for the modes $A_i(\vec x)$ is
a confining theory. However the latter magnetic  contributions start 
at order $g^6$,  and  cannot a priori be made
responsible for the poor apparent convergence that is seen up to order
$g^5$. 

We shall then focus in this paper on the soft contributions, i.e. 
those coming from the momentum scale $gT$. In principle
these can be calculated using perturbation theory, but as clear from 
(\ref{LQCDdr}) and (\ref{LQCDparam}) the corresponding
expansion parameter is $g_E^2/m_E\sim g$, so that the perturbative 
expansion converges only slowly, more slowly than the
perturbative expansion in the hard sector. As a minimal step towards 
a nonperturbative treatment of the soft sector, we
shall perform a simple loop expansion of (\ref{LQCDdr}), keeping the 
parameters $m_E$ and $g_E$ as given in terms of $g$
by the matching conditions, and not expanding them out in powers of 
$g$ in the final result. As we shall see, this simple
method leads to a significant improvement over the strict perturbative results.
We also consider the effect of 
including the four-loop logarithms recently obtained by Kajantie et al.,
and again find that strict perturbation theory has large scale
dependences which are drastically reduced when keeping
soft contributions untruncated.
For a particular choice of the 
sublogarithmic constant the untruncated four-loop
result is moreover close to the three-loop result, which itself agrees
well with available lattice results down to
temperatures of about $2.5T_c$.

In Sect.~\ref{sec:drspt} we consider a simple variational improvement of
perturbation theory in the form of dimensionally reduced
screened perturbation theory (DRSPT). This turns out to
be much simpler than HTLPT, while also allowing to resum
screening masses in a gauge invariant manner.
At two-loop order it leads to a result significantly
closer to lattice data than two-loop HTLPT, which suggests
that the partial failure of the latter as observed in
\cite{Andersen:2002ey,Andersen:2003zk} is due to
spurious hard contributions. At three-loop order, the gap
equation of DRSPT has no real solution, but this is
remedied upon inclusion of the four-loop logarithms.

In the final section we summarize and discuss our results
and try to put them into perspective with other techniques.

\section{Dimensional reduction beyond strict perturbation theory }

In the following we adhere to Ref.~\cite{Braaten:1996jr}
in the treatment of the dimensionally reduced effective theory
responsible for the contributions from the scale $gT$,
but we deviate  in that we shall not treat the soft sector strictly
perturbatively. We shall organize our
presentation by considering the successive approximations  obtained
by expanding the contributions of the hard modes to the pressure in 
powers of $\alpha_s=g^2/(4\pi)$. Each order in this
expansion defines also the accuracy with which the parameters of the 
effective Lagrangian are determined through
perturbative matching conditions.
However, at each level of approximation we shall
consider the effect of treating
the contributions of the soft sector
more completely by refraining from the specific truncations
usually employed in
a strictly perturbative expansion.

\subsection{One-loop order}

In massless QCD at one-loop order, the only contribution to the 
thermal pressure
(identical to minus the free energy)
  is coming predominantly from
hard momenta. Introducing a momentum cutoff $\Lambda_E$ to
separate the hard scale $2\pi T$ from the scale $gT$, one gets a 
contribution from
the soft sector  proportional to $T\Lambda_E^3$,
compensating a similar term in the interaction-free one-loop
contribution from hard modes. One ends up with the standard
ideal-gas result
$P_0=\pi^2T^4(8/45+7 N_f/60)$ for $N_f$ quarks and 3 colors.

Ref.~\cite{Braaten:1996jr} avoids the introduction of
momentum cutoffs by using dimensional regularization  for
the purpose of both ultraviolet and infrared regularizations.
Doing so, the one-loop result exclusively comes from the
hard modes: At this level of approximation where all interactions are 
neglected, the soft (zero) modes are
to be taken as massless, and their contribution
vanishes in dimensional regularization where scaleless
integrals are set to zero.

\subsection{Two-loop order}

The two-loop, i.e. order $\alpha_s$,  contribution of hard modes to 
the pressure is still
independent of a cutoff $\Lambda_E$, if this is handled by
dimensional regularization. It reads \cite{Shuryak:1978ut}
\be\label{P2}
P^{(2)}_{\rm hard}=-{2\pi\03}(1+{5\012}N_f)\alpha_s T^4.
\ee

At this level of approximation, the soft modes described by EQCD are 
massive, the mass being given by the
leading order Debye mass (\ref{LQCDparam}). The one-loop contribution 
to the pressure from EQCD is now, on
dimensional grounds, proportional to
$m_E^3$ times an overall factor of $T$. If only this contribution is 
added on to (\ref{P2}),
the result is the ill-behaved strictly-perturbative result to
order $g^3$ displayed in Fig.~\ref{fig:qcd}.
However, in the soft sector, we need not restrict ourselves to this 
one-loop approximation, but rather treat more
completely the interactions which are present in $\mathcal L_{EQCD}$. 
Thus, with the parameters of EQCD
determined by matching with the perturbative calculation at the 
present level of accuracy, we shall consider also the
two-loop contributions from the dimensionally reduced effective 
theory.  This yields the following contribution to the
pressure
\be\label{P12soft}
P^{(1)+(2)}_{\rm EQCD}
=-Tf_M^{(1)+(2)}
\ee
with $f_M^{(1)+(2)}$ given by \cite{Braaten:1996jr}
\be\label{fM12}
f_M^{(1)+(2)}=-{2\03\pi}m_E^3+{3\08\pi^2}\left(
{1\0\epsilon}+4\ln{\Lambda_E\02m_E}+3\right)g_E^2 m_E^2
+\delta f_E,
\ee
where $\Lambda_E$ is the scale of dimensional regularization in the 
soft sector, which may be loosely
associated with the separation
scale between the soft and the hard momenta.  Choosing the counterterm
$\delta f_E$ by minimal subtraction,
\be\label{dfE}
\delta f_E = - {3\08\pi^2}g_E^2 m_E^2 {1\0\epsilon},
\ee
leaves behind a dependence on
$\Lambda_E$.
Eventually, $\ln(\Lambda_E)$ has to combine with a
matching logarithm arising from the hard scales for
which $\Lambda_E$ provides the infrared cutoff. This
in fact happens after a careful perturbative matching
at three-loop order (see eq.~(\ref{P3h})).

Together (\ref{P2}) and (\ref{P12soft}) are accurate through
order $g^4\log(1/g)$, with an error of order $g^4$ as
to be expected from a two-loop calculation.
The coefficient of the $g^4\log(1/g)$ term is
correct provided\footnote{If $\Lambda_E$ is chosen to
be parametrically smaller
than $2\pi T$ by multiplying it by a fractional power $g^c$
with $0<c<1$, then the coefficient of $g^4\ln(1/g)$ is wrong
by a factor of $c$.}
  $\Lambda_E$ is set to a constant
times $T$. From its role in dimensional reduction,
it should be smaller than $2\pi T$ but larger than $gT$.
However, as remarked in Ref.~\cite{Braaten:1996jr}, the
introduction of cutoffs through dimensional regularization
leaves their relationship to momentum cutoffs undetermined,
and there could be a different relationship between
the scale of dimensional regularization and effective momentum
cutoffs depending on whether the latter act as IR or as UV
cutoffs. Ref.~\cite{Braaten:1996jr} even found that the
scale $\Lambda_E$ should be chosen {\em larger} than
the UV scale $\bar\mu\sim 2\pi T$ in order to
have optimal convergence of the hard contributions.
For simplicity, and to facilitate the comparison with other 
calculations where a similar identification is made,
we shall put
$\Lambda_E=\bar\mu$ in the following, i.e.~introduce only one scale 
for dimensional
regularization, and refrain from modifying it by hand depending
on whether the various logarithms can be identified as
arising from
regularization in the IR or in the UV.\footnote{The
ambiguity of the choice of $\Lambda_E$ can alternatively
be understood as arising from the freedom to renormalize
the effective 3-dimensional theory differently than by
minimal subtraction. Because in the next subsection
we shall compare with HTLPT
\cite{Andersen:1999fw,Andersen:1999sf,Andersen:2002ey,Andersen:2003zk}
where only minimal subtraction of additional divergences has
been considered, we stick to minimal subtraction in the following.
It should be kept in mind, however, that there is a source
of additional renormalization scheme dependence.}

\begin{figure}[h]
\includegraphics[bb=50 200 540 555,width=10cm]{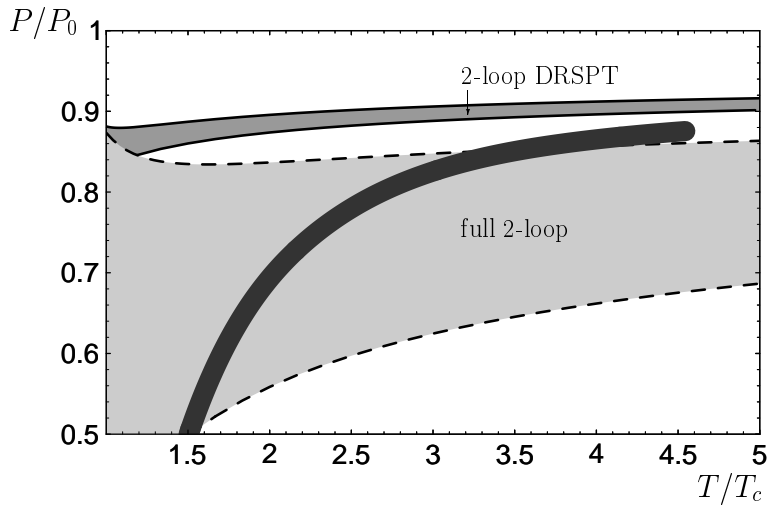}
\caption{\label{fig:2loop}
Two-loop pressure in pure-glue QCD with untruncated EQCD contributions
when $\bar\mu$ is varied between $\pi T$ and $4\pi T$ (broad gray band)
in comparison with the lattice result from Ref.~\protect\cite{Boyd:1996bx}
(thick dark-grey curve). The narrow darker-grey band above the former
is the result of 2-loop DRSPT considered in Sect.~\ref{sec:DRSPT2};
its lower boundary corresponds to
the extremal value when varying $\bar\mu$.}
\end{figure}

In Fig.~\ref{fig:2loop} we give the numerical evaluation of
the full 2-loop result obtained as indicated above,
for pure-glue QCD and $T$ between $T_c$ and $5T_c$
in analogy to
the strictly perturbative results of Fig.~\ref{fig:qcd}. 
We always use the standard\footnote{This is practically
indistinguishable from the full 2-loop solution as soon as
$T\gtrsim 2T_c$ (see Appendix of Ref.~\cite{McKeon:2002yx}).}
perturbative solution to
the two-loop renormalization group equation for $\alpha_s$
assuming $T_c/\Lambda_{\overline{\hbox{\scriptsize MS}}}=1.14$.
The UV renormalization scale $\bar\mu$ is varied about a
central value $2\pi T$ by a factor of 2, and it should be kept in 
mind that this variation also traces some of the
ambiguity in choosing $\Lambda_E$.
The resulting total scale dependence is comparable to
the scale dependence of the perturbative order-$g^4$ result. But
in contrast to both the result to order $g^3$, to which
it is perturbatively equivalent, and the order-$g^4$ result,
the untruncated 2-loop result
of dimensional reduction remains below the ideal-gas, and thus
has overlap with the lattice results,  which the former do not have.

Remarkably, the partial inclusion of order-$g^4$ effects arising
from a two-loop evaluation of EQCD is superior to a complete
order-$g^4$ evaluation in strict perturbation theory. Although
the former has an uncancelled $\Lambda_E$ dependence in addition
to the normal $\bar\mu$-dependence, 
numerically
the total scale dependence is not worse but even slightly better
than that of the perturbative order-$g^4$ result.

If we had not put $\lambda_E$ to zero on grounds that it starts
contributing only at order $g^6$, we would have obtained the
additional 2-loop term
\be\label{fM2lE}
f_M^{(2)}\Big|_{\lambda_E}={5\08\pi^2}m_E^2\lambda_E.
\ee
Inserting the leading-order value of $\lambda_E$, Eq.~(\ref{lambdaELO}),
this contribution is not only of order $g^6$, but it is also
numerically quite small in comparison with the other
two-loop contributions (\ref{fM12}) even when $g\sim 1$.

\subsection{Three-loop order}

The three-loop  contribution of the hard modes to the pressure is
no longer $\Lambda_E$-independent, because it has to be
matched with the minimally subtracted, and thus $\Lambda_E$-dependent,
EQCD theory. This has been calculated in \cite{Braaten:1996jr}
as\footnote{In Eq.~(54) of Ref.~\cite{Braaten:1996jr} the terms corresponding
to the second and fourth term on the r.h.s. of Eq.~(\protect\ref{P3h}) have
a different sign due to a typographical error.}
\bea\label{P3h}
P^{(3)}_{\rm hard}&=&{8\pi^2\045}T^4 \biggl\{
244.9+17.24N_f-0.415N_f^2
+{135}\left(1+{N_f\06}\right)\ln{\Lambda_E\02\pi T}\nn\\
&&\qquad\qquad-{165\08}\left(1+{5\012}N_f\right)\left(1-{2\033}N_f\right)
\ln{\bar\mu\02\pi T}
\biggr\}\left({\alpha_s\0\pi}\right)^2.
\eea
Similarly, the mass parameter of EQCD can be obtained by a matching
calculation of two-loop self-energies as \cite{Braaten:1996jr}
\bea\label{mE22}
m_E^2&=&(2\pi T)^2 {\alpha_s\0\pi} \biggl\{
\left(1+{N_f\06}\right) + {\alpha_s\04\pi} \biggl[
5+22\gamma+22\ln{\bar\mu\04\pi T} \nn\\&&\qquad\qquad
+{N_f\03} \left({1\02}-{8}\ln 2+{7}\gamma+{7}\ln{\bar\mu\04\pi T}\right)
+{N_f^2\09}\left(1-2\gamma-2\ln{\bar\mu\0\pi T}\right) \biggr]\biggr\}.
\eea

As for the  three-loop contribution from the soft sector, this is
given by the finite and thus $\Lambda_E$-independent expression
calculated in Ref.~\cite{Braaten:1996jr}
(neglecting $\lambda_E$-contributions now)
\be\label{fM3}
f_M^{(3)}={9\08\pi^3}\left({89\024}-{11\06}\ln2+{1\06}\pi^2\right)
g_E^4\,m_E^{\phantom4}.
\ee

\begin{figure}[h]
\includegraphics[bb=50 200 540 555,width=10cm]{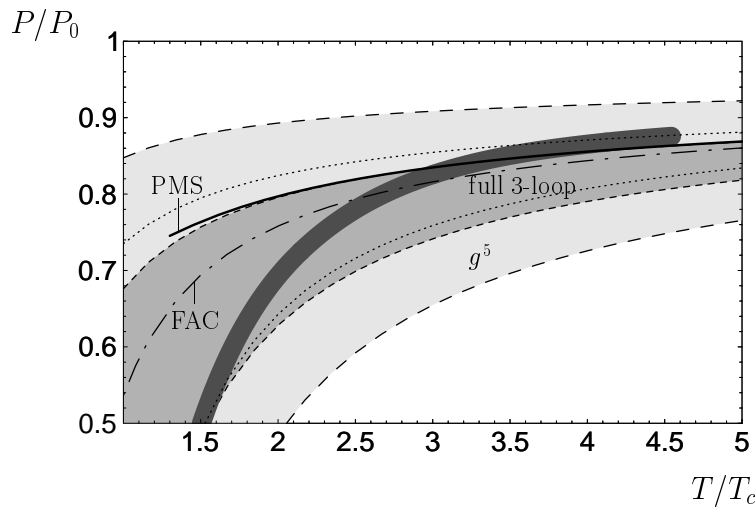}
\caption{\label{fig:3loop}
Three-loop pressure in pure-glue QCD with untruncated EQCD contributions
when $\bar\mu$ is varied between $\pi T$ and $4\pi T$ (medium-gray band); the
dotted lines indicate the position of this band when only the leading-order
result for $m_E$ is used.
The broad light-gray band underneath is the strictly perturbative result to
order $g^5$ with the same scale variations. The full line gives the
result upon extremalization (PMS) with respect to $\bar\mu$ (which does
not have solutions below $\sim 1.3T_c$); the dash-dotted line corresponds
to fastest apparent convergence (FAC) in $m_E^2$, which sets
$\bar\mu\approx 1.79\pi T$.}
\end{figure}

In a strictly perturbative evaluation which drops all
terms of order $g^6$, the sum of the
hard and soft contributions are $\Lambda_E$-independent
as they should be. That is, at order $g^4$, the term  $\propto 
\ln\Lambda_E$ in (\ref{P3h}) cancels against the
corresponding one in  (\ref{P12soft}). These perturbative 
contributions, evaluated numerically, give the result marked
``$g^5$'' in Fig.~\ref{fig:qcd} or Fig.~\ref{fig:3loop}.

However, our aim is to go beyond such perturbative results, and as a 
simple approximation in this direction, we
consider keeping the soft contributions (\ref{fM12}) and (\ref{fM3})
untruncated 
when the perturbatively determined
$m_E^2$ is inserted. This then corresponds to a selective
summation of higher-order effects that may help improve
the convergence of perturbation theory, although these higher
order terms are $\Lambda_E$-dependent.\footnote{Specifically, they
lead to a $g^6\ln(g)$ contribution with the constant under
the log carrying the ambiguity in $\Lambda_E$ if the latter
is proportional to $T$. Numerically, however, this $g^6\ln(g)$ contribution
is completely negligible compared to the
recently determined \cite{Kajantie:2002wa}
perturbative $g^6\ln(g)$ contribution appearing at 4-loop
order.} 

Setting $\Lambda_E=\bar\mu$ the ambiguity in choosing
$\Lambda_E$ contributes to the scale dependence and
is included in our error bands as
$\bar\mu$ is varied around its central value.\footnote{These error bands
would of course be widened by an independent variation, so the
scheme dependences displayed in our figures are certainly
underestimated to some extent.}
The result is shown in Fig.~\ref{fig:3loop}.
Despite the incomplete cancellation of $\Lambda_E$ in
the untruncated evaluation of the three-loop result, we observe
a considerable reduction of the
total scale dependence compared to strict
perturbation theory.

Furthermore, compared to the full two-loop result, the
three-loop result stays within the (rather large) uncertainties
of the former. While the uncertainties remain sizeable even
at three-loop order, the overall picture is a remarkable
improvement over strict perturbation theory, where the
results jump about the ideal-gas value and the renormalization
scale dependence {\em increases} steadily with the
highest power of $g$ reached.

It is interesting to observe that very similar results are obtained if
only the lowest order Debye mass (\ref{LQCDparam})
is used in this calculation, in place of the full order $g^4$ expression
(\ref{mE22}). This is displayed in Fig.~\ref{fig:3loop} by
the dotted lines.
It indicates that what matters
here is the accuracy with which one treats the soft
sector, more than the accuracy with which the
parameters of $\mathcal L_{EQCD}$ are determined.

Whereas in the untruncated two-loop result the scale dependence
is monotonic and does not allow for its elimination
by a principle of minimal sensitivity (PMS), such an
elimination turns out to be possible
in the three-loop result\footnote{This is possible in
fact only if the complete order-$g^4$ expression for the Debye
mass is used.}. Choosing $\alpha_s(\bar\mu)$ according to the 2-loop
renormalization group equation,
we find an extremum
of the untruncated three-loop result as a function of $\bar\mu$.
As shown in Fig. \ref{fig:3loop},
the corresponding pressure values are in fact remarkably close
to the lattice data for $T\gtrsim 3T_c$.

The extremum is only a local one with respect to $\bar\mu$.
For large temperatures $T\gtrsim 10T_c$ it
occurs at $\bar\mu\sim 2\pi T$, which following Ref.~\cite{Braaten:1996jr}
we have taken
as central value because it is the spacing of the Matsubara
frequencies. For smaller temperatures, the required value
of $\bar\mu$ increases and exceeds $4\pi T$ below $T\sim 3T_c$,
where the lattice data start to deviate from the three-loop
pressure. For still smaller temperatures $T\lesssim 2T_c$
the required $\bar\mu$ becomes unreasonably large, and finally
for $T\lesssim 1.3 T_c$ the local extremum
disappears completely.
(The strictly perturbative result, on the other hand,
has a monotonic, i.e.~run-away, scale dependence
for all $T$.)%

\begin{figure}[t]
\includegraphics[bb=50 200 540 555,width=10cm]{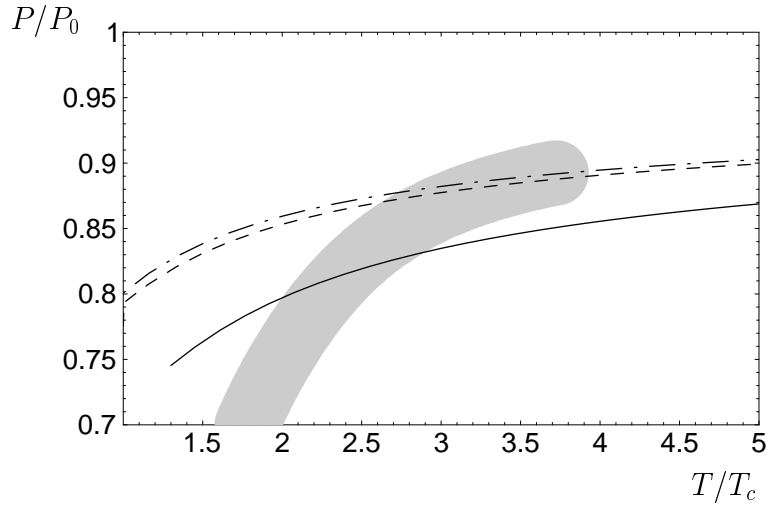}
\caption{\label{fig:nf023}
PMS-extremalized full-three-loop pressure in QCD with $N_f=0$ (full line),
$N_f=2$ (dashed line), and $N_f=3$ (dash-dotted line)
in comparison with the estimated continuum extrapolation
of QCD with 2 light quark flavors of Ref.~\cite{Karsch:1999vy}.}
\end{figure}

For completeness, we also give the numerical results obtained
by including $N_f=2$ and 3 massless quark flavors. The scale
selected by the minimal sensitivity turns out to be slightly
higher than at $N_f=0$, it also becomes large for $T\lesssim 2 T_c$,
but the extremum exists down to $T_c$ now. There exist no
reliable continuum extrapolated lattice data for this case yet,
but in Ref.~\cite{Karsch:1999vy} an {\em estimated} continuum extrapolation
has been given for $N_f=2$ light quark flavors. This is compared
with the extremalized full 3-loop results in Fig.~\ref{fig:nf023}.
The fact that the cases $N_f=2$ and 3 are nearly degenerate (when
normalized to the respective ideal-gas values and plotted as a
function of $T/T_c$ with the respective critical temperature [assumed
to be $1.14 \Lambda_{\overline{\hbox{\scriptsize MS}}}$])
is consistent with lattice results \cite{Karsch:1999vy}, and
is very similar to the pattern observed in the ``NLA'' results
of Ref.~\cite{Blaizot:2000fc} for the entropy.

\subsection{Four-loop order}

In an impressive four-loop calculation, the authors of 
Ref.~\cite{Kajantie:2002wa}
have recently determined the last 
coefficient in the perturbative expansion of thermal
pressure of QCD that can be computed analytically.
At four-loop order in the effective theory (\ref{LQCDdr})
there appear two logarithmic terms whose coefficents have
been obtained as \cite{Kajantie:2002wa}
\be\label{P4s}
P_{\rm soft}^{(4)}/T = N_g {(N g_E^2)^3 \0(4\pi)^4}
\left( \left[ {43\012}-{157\pi^2\0768} \right] \ln{\Lambda_E\0g_E^2}
+ \left[ {43\04}-{491\pi^2\0768} \right] \ln{\Lambda_E\0m_E}+c \right),
\ee
where the first logarithm is in fact from the magnetostatic sector.
To obtain the complete $g^6\ln g$-contribution in the QCD pressure
one now also needs $g_E^2$ to order $g^4$,
given in Ref.~\cite{Kajantie:1997tt} as (for $N_f=0$)
\be\label{gEcorr}
g_E^2=4\pi \alpha_s \left( 1 +
{\alpha_s\04\pi} \left[
22(\ln{\bar\mu\04\pi T}+\gamma)+1 \right] \right)
\ee
The constant $c$ in (\ref{P4s}), however, is strictly nonperturbative
and can in principle be determined by lattice simulations of
the dimensionally reduced theory, but require even more (also
analytical) work
\cite{Kajantie:2002wa}.

Also, the four-loop contribution to $P_{\rm hard}$ is not yet
known, but all the terms proportional to $g^6$ and involving explicit
logarithms of $\bar\mu$ or $\Lambda_E$ are
determined by the $\bar\mu$ and $\Lambda_E$
independence of the total pressure to order $g^6$, and have been
given explicitly in Ref.~\cite{Kajantie:2002wa}---only
a constant times $g^6$ thus remains undetermined.
Equating again $\Lambda_E$ with $\bar\mu$, the
four-loop contribution to be added to the above three-loop one when
evaluated with (\ref{gEcorr}) can be written as (again for $N_f=0$ only)
\bea\label{P4tot}
P^{(4)} &=& {8\pi^2\045}T^4 \biggl\{{21945\016}\ln^2{\bar\mu\02\pi T}
+2676.4 \ln{\bar\mu\02\pi T} \biggr\}\left(\alpha_s\0\pi\right)^3
\nonumber\\
&&+{27 \032\pi^4}
\left( \left[ {43\012}-{157\pi^2\0768} \right] \ln{T\0g_E^2}
+ \left[ {43\04}-{491\pi^2\0768} \right] \ln{T\0m_E}+ 7.57\,\delta\right)g_E^6T,
\eea
where for easier comparison with Ref.~\cite{Kajantie:2002wa}
we have collected all undetermined constants 
to this
order in a new constant $\delta$, 
chosen such that the $g_E^6T$ contribution in (\ref{P4tot}), 
when expanded in $g$, is proportional to $(\ln(1/g)+\delta)$.

The unknown coefficient $\delta$ of course leaves the numerical
outcome completely open until the required
analytical and numerical calculations will have
been performed. However, it has been observed in
Ref.~\cite{Kajantie:2002wa} that this coefficient could well be
such that the perturbative result follows closely the 4-d
lattice results. In obtaining numerical results, Ref.~\cite{Kajantie:2002wa}
in fact used a particular optimized renormalization scheme,
introduced in \cite{Kajantie:1997tt}, which also involves
keeping the parameters of the effective theory unexpanded.
(Refs.~\cite{Kajantie:2000iz,Kajantie:2002wa} also mentioned
that this reduces the scale dependence.)

\begin{figure}[h]
\includegraphics[bb=50 200 540 555,width=10cm]{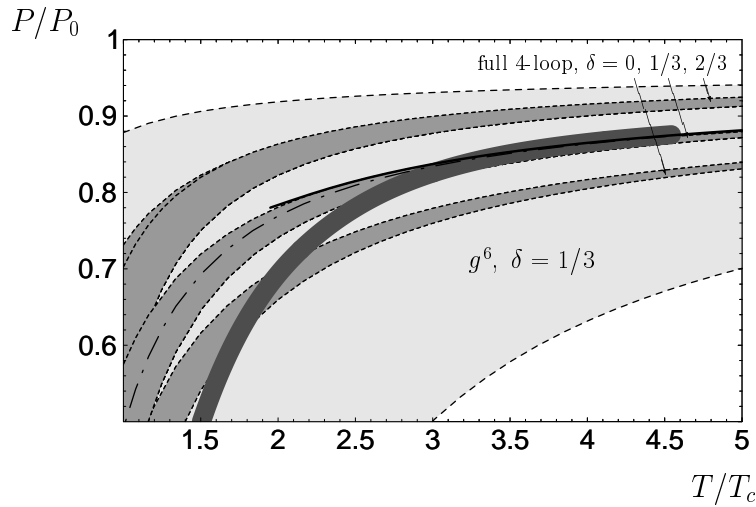}
\caption{\label{fig:4loop}
Four-loop pressure in pure-glue QCD 
including the recently determined $g^6\ln(1/g)$ contribution of
\cite{Kajantie:2002wa} together with three values for
the undetermined constant $\delta$ in Eq.~(\ref{P4tot})
when evaluated fully with $\bar\mu$ varied between $\pi T$ and
$4\pi T$ (medium-gray bands).
The broad light-gray band underneath is the strictly perturbative result to
order $g^6$ corresponding to the central value $\delta=1/3$.
The full line gives the untruncated
result with $\delta=1/3$ extremalized with respect to $\bar\mu$ (which does
not have solutions below $\sim 1.9T_c$); the dash-dotted line corresponds
to fastest apparent convergence (FAC) in $m_E^2$, which sets
$\bar\mu\approx 1.79\pi T$.}
\end{figure}

In Fig.~\ref{fig:4loop} we present our numerical results obtained by
adding the 4-loop terms (\ref{P4tot}) to the full 3-loop result 
of the previous section, but now evaluated with (\ref{gEcorr})
in all terms involving $g_E$,
for the possibilities $\delta=0,\,1/3,\,2/3$. This differs from 
\cite{Kajantie:2002wa}
in that we treat hard contributions strictly perturbatively and
only soft ones without truncations. Furthermore, we
use the perturbative 2-loop running coupling\footnote{%
Ref.~\cite{Kajantie:2002wa} found $\delta\approx 0.7$ to give
results which agree well with the 4-d lattice results. The difference
to our central value of 1/3 is mainly due to the fact that
we used 2-loop rather than 1-loop running coupling.
Like Ref.~\cite{Kajantie:2002wa} we included also the
contribution (\ref{fM2lE}), which is however fairly small.}
and vary $\bar\mu$ about $2\pi T$ by a factor of 2 as above.
The results, which are displayed by the 
medium-gray areas, show a remarkably small
scale variation. By contrast, the strictly perturbative evaluation
(given for $\delta=1/3$ only)
shows an {\em increased} scale dependence when compared to
the 3-loop results.

Also like in the previous 3-loop case, we find that the
untruncated result has again a nonmonotonic scale dependence
which makes it possible to fix the scale by PMS.\footnote{This
has been observed also before in Ref.~\cite{Laine:2003ay},
but using one-loop running and the particular
parametrization of Ref.~\cite{Kajantie:2002wa}.} The result
is again close to the FAC choice considered previously as well.
The strictly perturbative result on the other hand, has
a run-away scale dependence for almost all $T$.

\section{Screened perturbation theory in the soft sector}
\label{sec:drspt}

In massless scalar field theories, which have a poorly convergent
perturbative series for the thermal pressure similar to
what is found in QCD, Karsch, Patk\'os and Petretcky  and others
\cite{Karsch:1997gj,Chiku:1998kd,Andersen:2000yj,Andersen:2001ez}
have proposed to ameliorate the situation by a variationally
improved perturbation theory which uses a simple mass term
as variational parameter. In this so-called screened perturbation
theory (SPT) the mass term is part of the tree-level Lagrangian as well
as the interactions, where it is counted like a one-loop counterterm.

This approach has been extended to QCD
by Andersen et 
al.~\cite{Andersen:1999fw,Andersen:1999sf,Andersen:2002ey,Andersen:2003zk}
by using the hard-thermal-loop (HTL) action in place of a simple mass term,
turning its prefactor, which is proportional to the Debye mass squared,
into a variational parameter.
This HTL-screened perturbation theory (HTLPT) has been recently
carried through to two-loop order in QCD in
Refs.~\cite{Andersen:2002ey,Andersen:2003zk}.
The result is perturbatively correct to order $g^4\ln(g)$
and has been found to give rather stable results which are smaller
than the ideal-gas pressure, but significantly above the
lattice results.

In this section we investigate the possibility of improving
the soft contributions considered above to two and three loop
order by a dimensionally reduced screened perturbation theory (DRSPT)
for EQCD defined by trivially rewriting the EQCD Lagrangian according to
\be\label{EQCDSPT}
\mathcal L_{\rm EQCD}=
{1\04}F_{ij}^a F_{ij}^a + {1\02} (D_i A_0)^a (D_i A_0)^a
+{1\02}(m_E^2+\delta m^2) A_0^a A_0^a
-{1\02}\delta m^2 A_0^a A_0^a.
\ee
As above we take $m_E^2$ to be determined by perturbative matching
to order $g^2$ and $g^4$ when calculating the pressure to two and
three loop order, respectively, and we neglect $\lambda_E$ because,
as we have seen above,
it starts to contribute only at order $g^6$ with small
numerical effects.

\subsection{Two-loop order}
\label{sec:DRSPT2}

In DRSPT to two-loop order, the result for the pressure is
given by the sum of (\ref{P2}) and (\ref{P12soft}) except that
the latter now involves
\be\label{fM12SPT}
f_M^{(1)+(2)}=-{2\03\pi}m^3+{3\08\pi^2}\left(
{1\0\epsilon}+4\ln{\Lambda_E\02m}+3\right)g_E^2 m^2
+{m\0\pi}\delta m^2 + \delta f_E^{\rm DRSPT}.
\ee
where $m^2=m_E^2+\delta m^2$ according to (\ref{EQCDSPT}).
The term ${m\0\pi}\delta m^2$ arises
from a one-loop diagram where the counterterm $\delta m^2$, 
which itself has to be counted as a one-loop quantity, has been inserted.

SPT generally produces additional UV divergences and associated
scheme dependences, which can be seen here in that
the pole term in the second term
on the right-hand side of Eq.~(\ref{fM12SPT}) no longer matches (\ref{dfE}).
Following 
Refs.~\cite{Karsch:1997gj,Chiku:1998kd,Andersen:2000yj,Andersen:2001ez,Andersen:2002ey,Andersen:2003zk} 
we
treat those by minimal subtraction. 
This means that 
we introduce
a counterterm $\delta f_E^{\rm DRSPT}$
with $\delta f_E^{\rm DRSPT}-\delta f_E
\propto g_E^2 \delta m^2 {1\0\epsilon}$, which can be discarded
in the perturbative matching as being of higher order and will
disappear in fact at the next loop order (see below).
However, the replacement of $m_E^2$ by $m_E^2+\delta m^2$ has
the effect of modifying the dependence on $\Lambda_E$
at the present loop order.

The untruncated two-loop result, including now (\ref{fM12SPT})
in place of (\ref{fM12}), can be optimized by a variational
principle (principle of minimal sensitivity) for $\delta m^2$.
This leads to the variational mass
\be\label{mSPT2}
m^2=m_E^2-3\alpha_s(\bar\mu) T m\left(4\ln{\Lambda_E\02m}+1\right).
\ee
As before and similar to what is done in
Refs.~\cite{Andersen:2002ey,Andersen:2003zk} we equate $\Lambda_E$
and $\bar\mu$. Since the latter is always set proportional to $T$,
this gives rise to a term $\delta m^2\propto g^3T^2\ln(g)$, like
in the actual next-to-leading order Debye mass in QCD
\cite{Rebhan:1993az
},
however both coefficient and sign are different (and the
logarithm involves an UV rather than an IR cutoff).
This is no contradiction, however, since
the variational mass (like the effective mass parameter $m_E$)
is not identical with the Debye mass responsible for
exponential screening behavior
of the full electrostatic propagator. In fact, we shall see below 
that at higher loop orders the gap equation does
not contain a logarithmic term.

The simple ``gap equation'' (\ref{mSPT2}) is different from
the one obtained in 2-loop HTLPT \cite{Andersen:2002ey,Andersen:2003zk}.%
\footnote{When fermions are included in 2-loop HTLPT \cite{Andersen:2003zk}
the latter gives rise to a gap equation for fermions as well,
whereas in DRSPT fermions contribute only through the
parameter $m_E$.}
It is much simpler in form, and it turns out to have a numerical
solution that behaves differently from that of 2-loop HTLPT.
While the latter increases sharply at large coupling, the
solution of (\ref{mSPT2}) saturates 
at the value $m\approx 0.503 T$
as $\alpha_s$ increases when $N_f=0$ and $\Lambda_E=\bar\mu=2\pi T$
[for larger/smaller $\bar\mu$ the saturation occurs at smaller/larger values;
for $\bar\mu<1.4\pi T$ there is a maximum value of $\alpha_s$ beyond
which solutions no longer connect continuously to the perturbative
leading-order result---for instance, for our lowest value
$\bar\mu=\pi T$ this restricts $\alpha_s$ to less than $0.34$,
which however presents no problem to the following application].

\begin{figure}
\includegraphics[bb=50 200 540 555,width=10cm]{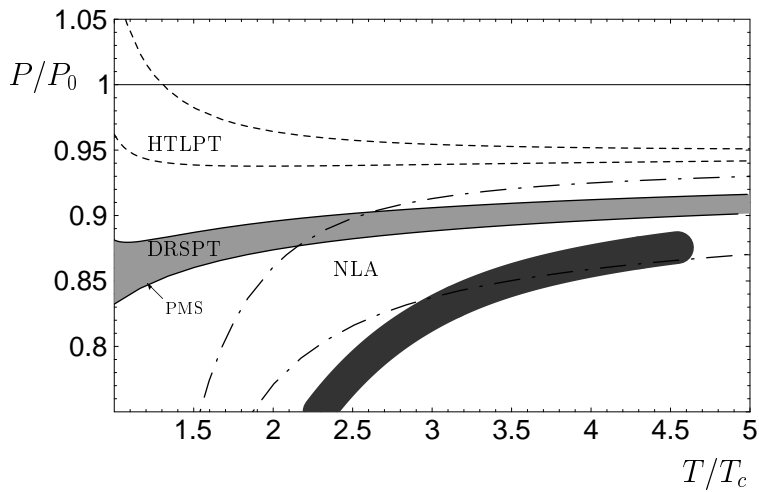}
\caption{\label{fig:drspt}
The two-loop DRSPT result for pure-glue QCD (gray band) in comparison with
the two-loop HTLPT result of Ref.~\cite{Andersen:2002ey} (dashed lines)
and the ``NLA''
result of Ref.~\cite{Blaizot:2000fc} (dash-dotted lines), all
with $\bar\mu$ varied around $2\pi T$ by a factor of 2.}
\end{figure}

The thermal pressure at two-loop DRSPT is obtained by evaluating
(\ref{fM12SPT}) at the solution of (\ref{mSPT2}) and
combining it with the hard contribution (\ref{P2}). The result
for pure-glue QCD is given in Fig.~\ref{fig:2loop} where it is
compared with the untruncated two-loop result which uses
the perturbative leading-order mass instead of the solution of
the gap equation (\ref{mSPT2}). In Fig.~\ref{fig:drspt} the
two-loop DRSPT result is furthermore compared to the
result of the two-loop HTLPT calculation of
Ref.~\cite{Andersen:2002ey} and the ``NLA''
result of Ref.~\cite{Blaizot:2000fc} which is
based on the $\Phi$-derivable two-loop expression for the entropy
evaluated with HTL propagators that include next-to-leading order
corrections to asymptotic thermal masses.
We find that the two-loop DRSPT result has
a rather small scale dependence like the
two-loop HTLPT result, but is significantly below
the latter and thus closer to the lattice data.
The nonlinear scale dependence of the DRSPT result in fact
makes it possible to eliminate the scale dependence by
a principle of minimal sensitivity. The result is given by the
lower boundary of the DRSPT band in Fig.~\ref{fig:drspt}.
This optimized DRSPT result
happens to lie right at the center of the estimated error
of the NLA result of Ref.~\cite{Blaizot:2000fc} for $T>3T_c$
[the fact that the latter sharply drops close to $T_c$ is in fact not
a prediction of the NLA result but comes from the necessity
to fix an integration constant which has been chosen such
that $P(T_c)=0$].

Here we have neglected the soft 2-loop contribution involving $\lambda_E$
of Eq.~(\ref{fM2lE}) because HTLPT does not include a comparable term.
When (\ref{fM2lE}) is included, the 2-loop gap equation is modified
by an extra contribution $-{5\03\pi}(9-N_f)\alpha_s^2 T m$ on the
right-hand side of Eq.~(\ref{mSPT2}). 
Its (rather small) effect is displayed on the occasion of
the comparison with 4-loop DRSPT in Fig.~\ref{fig:4lspt}.

\subsection{Three-loop order}

The three-loop free energy of EQCD in DRSPT can be easily
derived from the results of Ref.~\cite{Braaten:1996jr}
\bea\label{fM123SPT}
f_M^{(1)+(2)+(3)}&=&-{2\03\pi}m^3+{3\08\pi^2}\left(
{1\0\epsilon}+4\ln{\Lambda_E\02m}+3\right)g_E^2 m^2_E
+{9\08\pi^3}\left({89\024}-{11\06}\ln2+{1\06}\pi^2\right)
g_E^4\,m \nn\\ &&
+{m\0\pi}\delta m^2
+{3\04\pi^2}g_E^2 \delta m^2
-{1\04\pi m}(\delta m^2)^2.
\eea
where $m^2=m_E^2+\delta m^2$ is now defined with
the order-$g^4$ result for $m_E^2$, Eq.~(\ref{mE22}), and
where up to two insertions of the SPT counterterm $\delta m^2$
have to be included.

At this order the $\delta m^2$ counterterm of DRSPT restores
the UV divergent part to be proportional to $m_E^2$
(second term on the r.h.s.\ of Eq.~(\ref{fM123SPT})), and because
no further pole terms appear at three-loop order, the additional
UV divergences of SPT drop out altogether.

The attempt to determine $m$ by a variational principle leads to
the simple quartic gap equation
\be\label{mSPT3}
G^{(3)}(m)\equiv
{1\08}(m^2-m_E^2)^2+{g_E^2N_c\04\pi} (m^2-m_E^2)\,m+
\left({g_E^2N_c\04\pi}\right)^2
\left({89\024}-{11\06}\ln2+{1\06}\pi^2\right)m^2 =0\,,
\ee
which we have written out for general color number $N_c$.
However, one can readily prove that this equation has no solution
that connects continuously to the perturbative result $m_E$
as $\alpha_s\to 0$ (for any value of $N_c$; also
inclusion of the term (\ref{fM2lE}) does not change this situation).

The same problem in fact occurs in SPT in scalar $\phi^4$ theory
to three-loop order \cite{Andersen:2000yj}. Like
Ref.~\cite{Andersen:2000yj} one may look
for alternative prescriptions to set up a gap equation.
The simplest conceivable option, however, clearly is to set $\delta m^2=0$.
This trivially connects to perturbation theory, and amounts
to our previous suggestion of keeping the 3-loop contributions
of EQCD untruncated, which is in fact most natural since $m_E$
in EQCD is just a mass parameter. Fortunately, the now nonlinear
scale dependences can be eliminated by a principle of minimal
sensitivity, as we have discussed above, so in a sense our
above improvement may be taken as a trivial implementation of
DRSPT,
with the variational parameter being the renormalization
and separation scale rather than an additional mass term.

\subsection{Four-loop order}

In DRSPT to four-loop order (neglecting $\lambda_E$ contributions)
we have
\bea\label{f4spt}
f_M^{(4)} &=& -{27 g_E^6 \032\pi^4}
\left( \left[ {43\012}-{157\pi^2\0768} \right] \ln{\Lambda_E\0g_E^2}
+ \left[ {43\04}-{491\pi^2\0768} \right] \ln{\Lambda_E\0m}+c \right)\nn\\
&&-{9g_E^4\016\pi^3}\left({89\024}-{11\06}\ln2+{1\06}\pi^2\right)
{\delta m^2\0m}-{3g_E^2\08\pi^2}{(\delta m^2)^2\0m^2}
-{(\delta m^2)^3\024\pi m^3},
\eea
which is to be added to (\ref{fM123SPT}), evaluated now also
with $g_E^2$ to order $g^4$ as given by (\ref{gEcorr}).

The variational gap equation associated with the sum of (\ref{fM123SPT})
and (\ref{f4spt}) does not involve the unknown constant $c$ and
is therefore known completely to order $g_E^6$ (in fact,
the contribution (\ref{fM2lE}) proportional to $\lambda_E$ does
not enter either, while $\lambda_E^2$ contributions to (\ref{f4spt})
are already of order $g^9$ or higher). This equation is now
of sixth order in $m$ and reads
\be\label{mSPT4}
G^{(4)}(m)=(m^2-m_E^2)\,G^{(3)}(m)
+2\left({g_E^2N_c\04\pi}\right)^3\left[ {43\04}-{491\pi^2\0768} \right]
m^3=0.
\ee
As it turns out, the inclusion of the 4-loop logarithm lifts the
impasse encountered with DRSPT at 3-loop order: there is now
a (unique) solution to the 4-loop gap equation which does connect
continuously to perturbation theory and which 
is given by the quadratic gap equation\footnote{The numerical
coefficient therein is given by the (real) root of a cubic
equation involving the somewhat unwieldy constants appearing
in (\ref{fM123SPT}) and (\ref{f4spt}) and could in principle
be given in closed (but lengthy) form. 
}
\be\label{m4lg}
m^2=m_E^2 - 0.33808\,g_E^2 N_c m,
\ee
which appears as a factor of $G^{(4)}(m)$ and whose only solution with
real and positive $m$ is given by
\be\label{mSPT4n}
m=\sqrt{m_E^2+(0.16904\, g_E^2 N_c)^2}-0.16904\, g_E^2 N_c.
\ee

It is intriguing that the quadratic gap equation (\ref{m4lg}) is
of the same form as the one adopted in the NLA approximation
of Ref.~\cite{Blaizot:2000fc} for the asymptotic thermal masses,
and also the coefficients in the two gap equations happen
to be very close ($0.338N_c$
versus $N_c/\pi$). However, it should be emphasized that the
gap equation of DRSPT has no physical meaning outside of DRSPT.
Indeed there is no reason to expect the leading correction to
$m_E$ as prescribed by the 4-loop
DRSPT gap equation to remain the same at 5-loop level, 
if at that order solutions exist at all. At any given
loop order, the deviation of $m$ from $m_E$ only influences
the orders beyond the perturbative accuracy.

\begin{figure}
\includegraphics[bb=50 200 540 555,width=10cm]{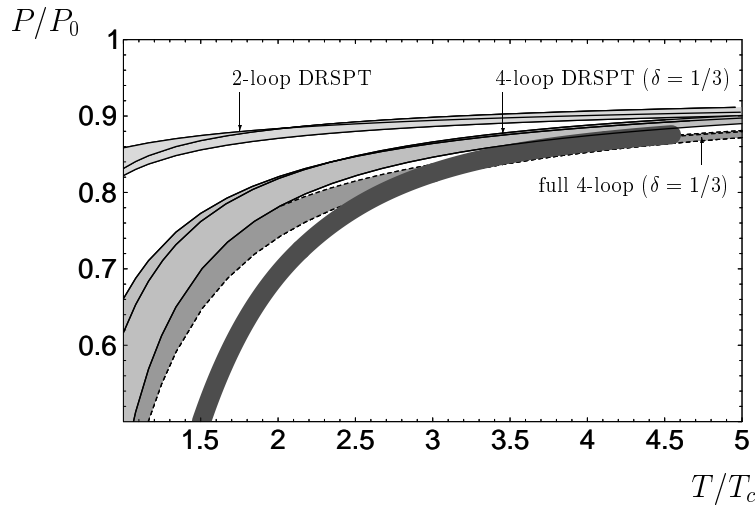}
\caption{\label{fig:4lspt}
The 4-loop DRSPT result for pure-glue QCD (3 full lines corresponding
to $\bar\mu=\pi T,2\pi T,4\pi T$ and $\delta=1/3$) in comparison
with the two-loop DRSPT result (now evaluated with
the contribution (\ref{fM2lE}) included) and the untruncated 4-loop with
$\delta=1/3$.}
\end{figure}

Numerically, the deviation of $m$ as given by (\ref{m4lg}) from
the perturbative value $m_E$ has some effect as displayed in
Fig.~\ref{fig:4lspt}. It turns out that the 4-loop DRSPT differs
from the untruncated 4-loop result of the previous section
in that it has a significantly larger $g^7$ coefficient
(by a factor of almost 6 when $\bar\mu=2\pi T$).
As a consequence, this gives a slightly different ``prediction'' for the
unknown $g^6$ constant $\delta$, but otherwise the
results are quite similar to those
obtained in the simple untruncated evaluation.

In fact, at 4-loop order the order-$g^7$ coefficient could in principle
be calculated completely, if $m_E^2$ is determined to 3-loop
accuracy and relevant higher-dimension operators in the
effective theory are included. (The order $g^6$ coefficient of course
remains beyond the reach of perturbation theory.)
In this case, however, 4-loop DRSPT
would spoil the achieved perturbative accuracy, because it
changes the $g^7$ coefficient without correcting these
changes through the SPT counter\-terms, which only take care of orders
up to and including $m^3T[(\delta m^2)/m^2]^3 \sim g^6 T^4$. 
Thus, beginning at
4-loop order, (DR)SPT ceases to be a possible improvement over (truncated
or untruncated) perturbation theory.


\section{Large-$N_f$ limit}

We finally consider also the recently solved
large-$N_f$ limit of QCD \cite{Moore:2002md,Ipp:2003zr,Ipp:2003jy}
which has been proposed as a testing ground for
improvements of perturbation theory.
In this limit only terms involving products $\alpha_s N_f$
are kept in the above results and $\alpha_s$ itself is taken to zero.
The dimensionally reduced theory is therefore non-interacting
in the large-$N_f$ limit and the soft contributions
are given exactly by $f_M=-{2\03\pi}m_E^3$. Still, at a given loop order
for the hard contributions, we can
investigate the difference between a strictly perturbative
evaluation of $f_M$ versus an untruncated one which resums
an infinite number of terms with odd powers in $g$.
Also
the gap equations of DRSPT become
trivial (but solvable): they all amount to setting $m^2=m_E^2$.

\begin{figure}
\includegraphics[bb=50 200 540 555,width=10cm]{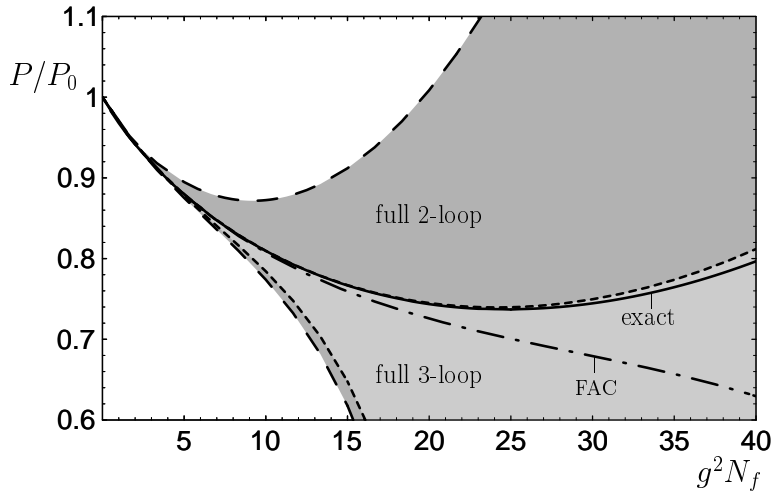}
\caption{\label{fig:lnf}
The exact result of the thermal pressure in the limit of large $N_f$
from Ref.~\cite{Ipp:2003zr},
normalized to that
of free gluons and as a function
of $g^2N_f(\bar\mu=\pi e^{-\gamma}T)$
(full line), in comparison with the untruncated
two-loop and three-loop results
(darker and lighter gray areas, respectively), each with
$\bar\mu$ varied
around a central value of $2\pi T$ by a factor of $e$.
The dash-dotted line corresponds to fastest apparent convergence.
}
\end{figure}

The results of a numerical evaluation of the untruncated two-loop
and three-loop results are compared with the exact result
of Ref.~\cite{Ipp:2003zr} in Fig.~\ref{fig:lnf}.
Again one can observe a great reduction of the scale dependence
by going from two-loop to three-loop order.
Compared to the strictly perturbative result to order $g^5$
(not displayed in Fig.~\ref{fig:lnf})
the reduction of the size of the scheme dependence is
less important than in the pure-glue case,
e.g., at $g^2N_f=10$
the reduction is about 16\%.

Also in contrast to the pure-glue case, the three-loop result
has a monotonic scale dependence, so the scale cannot be
fixed by minimal sensitivity.
As remarked in Ref.~\cite{Ipp:2003zr}, $\bar\mu$
could instead be fixed by fastest apparent convergence (FAC).
Requiring e.g.~that the $\alpha_s^2$ term in $m_E^2$ vanishes
leads to the choice $\bar\mu/T=\pi e^{{1\02}-\gamma}\approx 0.93\pi T$.
For this value the untruncated 3-loop result coincides with the
result to order $g^5$ in strict perturbation theory,
which in turn agrees quite well with the exact result
up to rather large coupling \cite{Ipp:2003zr}. So while this comparison
does not favor one over the other, it shows that with an optimal
choice of the renormalization scale both the perturbative result to
order $g^5$ and the untruncated 3-loop result fare rather well
in the large-$N_f$ limit.

If one applies the same prescription to the 3-loop result of
EQCD in pure-glue QCD, one is lead to setting
$\bar\mu/T=4\pi e^{-\gamma-5/22}\approx 1.79\pi T$. This
is lower but close to the scale selected by minimal sensitivity
of the untruncated 3-loop result. Correspondingly, the numerical result
following from the FAC choice
is fairly close to the
one obtained from minimal sensitivity (dash-dotted and full line
in Fig.~\ref{fig:3loop}, respectively), at least for $T\gtrsim 3T_c$, which
appears to validate the PMS results.\footnote{The upper boundary of the
range of the three-loop results shown in Fig.~\protect\ref{fig:lnf}
follows the exact result to even much larger values. There is
in fact a choice of $\bar\mu\approx 0.75 \pi T$ for
both the untruncated three-loop result and the strictly
perturbative result, where they become almost indistinguishable
from the exact result. A renormalization scale close to this
turns out to be favoured by applying FAC to the perturbative
result to order $g^6$, which
has been extracted numerically in the large-$N_f$ limit \cite{Ipp:2003jy},
but which is strictly non-perturbative in real QCD.}

\section{Discussion and Conclusion}

We have found that the scale dependence and convergence of
the results for the thermal pressure from perturbative QCD at
high temperature are significantly improved when the contributions
from soft scales as given by the effective dimensionally reduced theory
EQCD are not treated in strict perturbation theory.

In
particular, we have explored the predictions of a
simple loop expansion of EQCD, in which,
after strictly perturbative matching of the parameters of the
effective Lagrangian,
the results are not subsequently expanded out in powers of
$g$ and truncated. The result obtained at  the
two-loop level in this scheme includes contributions to order
$g^4\ln g$ completely
while being incomplete to order $g^4$, and is such that the
pressure no longer exceeds the
ideal-gas limit. The scale dependence is large, however the 
three-loop result is within the estimated
boundaries of the two-loop result.
This three-loop result has a smaller scale dependence than
that of strict perturbation theory to order $g^5$, and moreover
the scale dependence is nonmonotonic so that it can be
eliminated by a principle
of minimal sensitivity (PMS). The correspondingly optimized
result is rather close to the lattice data on
the continuum limit of pure-glue QCD for $T\gtrsim 3T_c$.
When including 4-loop effects, in particular the recently
determined $g^6\ln g$ contribution of Ref.~\cite{Kajantie:2002wa},
we find that this trend continues
and, in line with \cite{Kajantie:2002wa,Laine:2003ay},
that it is quite possible that all higher-order contributions
add up to a very small correction above $\sim 3 T_c$.

We have also considered variationally improved ``screened'' perturbation
theory in the dimensionally reduced theory (DRSPT), where it is a much
simpler, gauge invariant alternative to HTLPT (though not
extensible to dynamic quantities, as HTLPT in principle is).
The result for the pure-glue pressure when improved through 2-loop DRSPT
turns out to be significantly lower than that of HTLPT and
for $T\gtrsim 3T_c$ fairly close to the lattice results
as well as to the results of Ref.~\cite{Blaizot:2000fc}.

An obvious advantage of DRSPT over HTLPT is that it does not
modify the theory at hard momentum scales, where the HTL approximation
in general breaks down. The latter continues to be a good approximation
at hard momentum scales only at soft virtuality. In the entropy-based
HTL-resummations of Ref.~\cite{Blaizot:2000fc} it turns out that
the contributions are predominantly coming from hard momenta close
to the light-cone. 
In the HTLPT approach, on the other hand, spurious contributions
at hard momenta occuring at a given loop order are corrected
for by the specific counterterms of HTLPT at next loop order, so this may
present a problem at low loop orders. Our two-loop
DRSPT results seem to indicate that this is indeed the
reason for the difficulties of two-loop HTLPT
\cite{Andersen:2002ey,Andersen:2003zk}.

An unsatisfactory feature of DRSPT, as observed before in SPT
applied to scalar field theories and thus not unlikely to affect
HTLPT as well, is that at three-loop order the mass gap equation
does not have solutions which connect to perturbation theory.
This impasse happens to be lifted by the inclusion of
soft four-loop logarithms, and the result is then close
to that obtained by a simple untruncated evaluation of
all soft contribtions. Nevertheless, the gap equations of
(DR)SPT have no particular physical interpretation
(as discussed after Eqs.~(\ref{mSPT2}) and (\ref{mSPT4n})),
which casts some doubt on the systematics of SPT and its
usefulness in improving perturbation theory.

Evidently, our main result is that the convergence behavior of 
successive approximations to the pressure is dramatically
improved by abandoning strict
perturbation theory in the soft sector. Treating this sector beyond
strict perturbation theory is in
fact closer in spirit to the so-called $\Phi$-derivable 
approximations \cite{Baym:1962} which are the basis
for the resummation techniques developed in
Ref.~\cite{Blaizot:1999ip,
Blaizot:2000fc}.
Such approximations, when implemented in the soft sector,  may 
represent an interesting alternative to DRSPT.
DRSPT has a single variational parameter, the mass of electrostatic 
gluons. While this
has the advantage of great simplicity as well as gauge invariance,
the full self-energy of electrostatic gluons is a nonlocal
quantity. One might consider a $\Phi$-derivable approach
  which does not have the need for the specific counterterms
of SPT and try to construct improved approximations, which
are in principle gauge dependent, though such gauge dependences
are strongly suppressed at the variational point \cite{Arrizabalaga:2002hn}.

It would be interesting to compare such a dimensionally reduced
$\Phi$-derivable approach 
with the one based on the entropy formalism of Ref.~%
\cite{Blaizot:1999ip,
Blaizot:2000fc}.
In the latter, the emphasis is on dynamical quasiparticles,
which at two-loop order are interaction free. It should be
noted that this approach is dependent on a real-time formalism
which does not lend itself to dimensional reduction. Indeed,
it involves differentiating thermal distribution functions
at the stationary point, where the temperature-dependence of
spectral functions drops out. However, the relevant theory for the
soft modes (including hard ones with soft
virtuality) is known: at leading order this involves
the non-local hard-thermal-loop effective action
\cite{Braaten:1992gm,Frenkel:1992ts,Blaizot:1994be}).
In fact, the real-time approach might have advantages when it
comes to including the effects of high chemical potentials $\mu$.
In this case dimensional reduction does not occur. The
quasiparticle approach on the other hand
appears promising for covering the thermodynamics in
the entire $T$-$\mu$-plane
\cite{Peshier:1999ww,
Romatschke:2002pb,
Szabo:2003kg}.

We intend to investigate these matters in future work. From the
present study we conclude that perturbative QCD at high
temperature is not limited to $T\gg 10^5 T_c$
as previously thought \cite{Arnold:1995eb,Braaten:1996ju}
but, when supplemented by appropriate
resummation techniques for soft physics,
seems to be capable of remarkably good
quantitative predictions down
to $T\sim 2.5 T_c$.

\acknowledgments

We would like to thank Keijo Kajantie and Mikko Laine for
helpful correspondence and for suggesting to consider also
the 4-loop logarithms calculated in Ref.~\cite{Kajantie:2002wa}.
This work has been supported by the Austrian-French
scientific exchange program Amadeus of APAPE and \"OAD,
and the Austrian Science Foundation FWF, project no.~P14632.


\end{document}